\documentclass[12pt,draftcls,onecolumn]{IEEEtran}
\usepackage{cite}
\usepackage{amsmath,amssymb,amsfonts}
\usepackage{algorithmic}
\usepackage{graphicx}
\usepackage{textcomp}
\usepackage{multirow}
\usepackage{colortbl,booktabs}
\usepackage{subfig}
\newtheorem{definition}{Definition}
\newtheorem{lemma}{Lemma}
\newtheorem{theorem}{Theorem}
\newtheorem{assumption}{Assumption}
\newtheorem{corollary}{Corollary}
\def\BibTeX{{\rm B\kern-.05em{\sc i\kern-.025em b}\kern-.08em
    T\kern-.1667em\lower.7ex\hbox{E}\kern-.125emX}}
\begin{document}
\title{State-action control barrier functions: Imposing safety on learning-based control with low online computational costs}
\author{Kanghui He, Shengling Shi, \IEEEmembership{Member, IEEE}, Ton van den Boom, and Bart De Schutter, \IEEEmembership{Fellow, IEEE}
	\thanks{This paper is part of a project that has received funding from the European Research Council (ERC) under the European Union’s Horizon 2020 research and innovation programme (Grant agreement No. 101018826 - CLariNet). }
	\thanks{Kanghui He ({\tt\small k.he@tudelft.nl}), Ton van den Boom ({\tt\small a.j.j.vandenBoom@tudelft.nl}), and  Bart De Schutter ({\tt\small b.deschutter@tudelft.nl}) are with Delft Center for Systems and Control, Delft University of Technology, Delft, The Netherlands.}%
	\thanks{Shengling Shi ({\tt\small slshi@mit.edu}) is with the Department of Chemical Engineering, Massachusetts Institute of Technology, Cambridge, Massachusetts, USA.}}
\maketitle

\begin{abstract}
  Learning-based control with safety guarantees usually requires real-time safety certification and modifications of possibly unsafe learning-based policies. The control barrier function (CBF) method uses a safety filter containing a constrained optimization problem to produce safe policies. However, finding a valid CBF for a general nonlinear system requires a complex function parameterization, which in general makes the policy optimization problem difficult to solve in real time. For nonlinear systems with nonlinear state constraints, this paper proposes the novel concept of state-action CBFs (SACBFs), which do not only characterize the safety at each state but also evaluate the control inputs taken at each state. SACBFs, in contrast to CBFs, enable a flexible parameterization, resulting in a safety filter that involves a convex quadratic optimization problem, which significantly alleviates the online computational burden. We propose a learning-based approach to synthesize SACBFs. The effect of learning errors on the effectiveness of SACBFs is addressed by constraint tightening and introducing a new concept called contractive-set CBFs. This ensures formal safety guarantees for the learned CBFs and control policies. Simulation results on an inverted pendulum with elastic walls validate the proposed CBFs in terms of constraint satisfaction and CPU time.
\end{abstract}


\section{Introduction}
Learning-based control methods have demonstrated extensive success across many applications, such as autonomous vehicles \cite{fisac2018general} and robotics \cite{dawson2023safe}. Both supervised learning and reinforcement learning (RL) have been widely used for controller synthesis. However, learning-based controllers may provide unsafe control actions that result in undesirable or even destructive effects on the system. In control systems, safety means that the trajectories of the closed-loop system should satisfy state and input constraints for the entirety of the system’s evolution. The lack of safety guarantees limits the ability of learning-based control to achieve safety-critical tasks.

\textcolor{blue}{There has been an increasing interest in designing controllers for safety-critical systems. Safe RL is a subfield of RL focused on ensuring safety during or after the learning phase. Techniques in safe RL include penalty methods \cite{he2023approximate}, primal-dual policy optimization \cite{paternain2019learning}, constraint elimination \cite{zheng2021safe}, invariant sets \cite{li2022robust}, Lyapunov functions \cite{chow2018lyapunov}, and control barrier functions (CBFs) \cite{dawson2023safe,ames2016control}. These methods aim to prevent the agent from taking unsafe actions by either penalizing such actions in the reward function or by modifying the policy to avoid them. However, penalty-based approaches usually faces trade-offs between ensuring safety and achieving optimal performance, while policy refinement to ensure safety typically requires formal safety certificates, which can be challenging to obtain \cite{wabersich2023data}.} 

\textcolor{blue}{Model Predictive Control (MPC) is another approach extensively used to ensure safety in control systems \cite{rawlings2017model}. However, traditional online MPC can be computationally intensive for nonlinear constrained systems \cite{schwenzer2021review}. Using supervised learning to approximate the MPC policy is an emerging active research topic that can achieve satisfactory performance in both safety and online computational efficiency \cite{hertneck2018learning}.}

\textcolor{blue}{Ensuring long-term safety typically requires either long-horizon prediction \cite{didier2023approximate} or the computation of invariant sets \cite{li2022robust}. CBFs offer an alternative approach. As CBFs are energy-based functions, their sub-level sets can be used to characterize the long-term safety of a dynamical system, thereby avoiding the complexity associated with invariant-set representations and long-horizon predictions.} Given any learning-based controller without safety considerations, an online optimization problem, called a safety filter, is then solved to find the closest safe input to this controller, while safety is gained by adding a constraint based on a CBF into the optimization problem. \textcolor{blue}{The safety filter offers a modular approach to impose safety for any pre-existing controller with formal guarantees, and is more online computationally efficient than some other controllers (such as MPC) that need long-term prediction.} \textcolor{blue}{CBFs provide a theoretically sound way to ensure that long-term safety constraints are respected by a single CBF constraint, while safety filters adjust control inputs in real time to enforce compliance with the CBF constraint.} \textcolor{blue}{In addition to CBF-based safety filters, there are also predictive safety filters \cite{wabersich2023data} and safety filters that involve projection onto invariant sets \cite{karg2020efficient}. In this paper, we will use the term “safety filter” specifically to refer to a “CBF-based safety filter”.} 

\textcolor{blue}{In the realm of CBF-based methodologies, even though the state and input constraints are known, there is a lack of universally applicable methods for generating valid CBFs, necessitating reliance on manually designed or problem-specific CBFs.} For some particular kinds of systems such as linear, piecewise affine \cite{lazar2006stabilizing}, or polynomial systems \cite{prajna2007framework}, CBFs can be computed by solving convex optimization problems. For nonlinear systems with general constraints, using learning-based algorithms accompanied by advanced function approximators to estimate CBFs has been explored in several contexts. Roughly, there are four kinds of methods to learn CBFs: the optimization-based method \cite{robey2020learning}, the learner-verifier method \cite{dai2020counter}, the Hamilton-Jacobi (HJ) reachability method \cite{choi2021robust}, and the predictive safety filter method \cite{didier2023approximate,wabersich2022predictive}. The first two methods can be conservative, i.e., the learned safe set can be a small subset of the maximal controlled invariant set. In comparison, the HJ reachability method relies on the dynamic programming principle to approximate the maximal controlled invariant set iteratively, while the predictive safety filter method is based on a receding-horizon open-loop optimal control problem to implicitly determine the safe set, which also converges to the maximal controlled invariant set as the horizon goes to infinity \cite{korda2020computing}. A comprehensive comparison of HJ reachability, predictive safety filters, and CBFs can be found in \cite{wabersich2023data}.

No matter how the CBF is learned, to control a given system with safety guarantees, it is inevitable to solve an online optimization problem with CBF-based constraints, which is usually non-convex for discrete-time nonlinear systems \cite{agrawal2017discrete}. To obtain a satisfactory approximation accuracy, sophisticated function approximators such as deep neural networks (NNs) are commonly used to parameterize the CBF \cite{didier2023approximate,dawson2023safe}. However, this will inherently cause non-convexity and increase the complexity of the optimization problem. As a result, the increased online computational load makes the CBF-based approach unsuitable for situations where fast computation of control inputs is required. Besides, the effect of approximation errors on the validity of the learned CBFs and the resulting control policies has not yet been addressed, as highlighted in \cite{dawson2023safe}. \textcolor{blue}{Note that \cite{choi2021robust} introduces control barrier-value functions by unifying Hamilton-Jacobi (HJ) reachability and CBFs to address issues of finding valid CBFs. However, \cite{choi2021robust} focuses on theoretical guarantees of robust safety to disturbances but does not address learning errors.}

\textcolor{blue}{The current paper has the following main contributions:}

\noindent\textcolor{blue}{\textbf{(i) Computationally efficient safety enforcement using state-action CBFs:} We introduce the novel concept of state-action CBFs (SACBFs). Unlike standard CBFs \cite{choi2021robust,anand2021safe,agrawal2017discrete}, which are often computationally intensive when ensuring safety in nonlinear systems, SACBFs provide a flexible parametrization.The flexibility lies in the ability to enforce safety constraints through a convex quadratic optimization problem, significantly reducing the computational burden during online operations.}

\noindent\textcolor{blue}{\textbf{(ii) Handling approximation errors in learned CBFs:} We propose a constraint tightening approach alongside the novel concept of contractive-set CBFs to address approximation errors in learned CBFs. This ensures that the invariance property of CBFs is maintained even when approximation errors occur, which is not yet achieved in existing work on learning CBFs \cite{choi2021robust,robey2020learning,didier2023approximate,srinivasan2020synthesis}. Additionally, we examine the relationship between SACBFs and contractive-set CBFs, and develop a new learning-based method to approximate SACBFs. This ensures the safety of the policy filtered by SACBFs, provided the approximation error is sufficiently small. Besides, we discuss the trade-off between online computational efficiency and safety when learning the proposed SACBFs.}

\section{Preliminaries and problem formulation} 

\subsection{Preliminaries}
We consider a deterministic discrete-time nonlinear system
\begin{equation}\label{system}
	x_{t+1}=f\left(x_t, u_t\right), \quad t=0,1, \ldots,
\end{equation}
where $x_t \in \mathcal{X} \subseteq \mathbb{R}^{n_x}$ and $u_t \in \mathcal{U}\subseteq \mathbb{R}^{n_u}$ are the state and the input at time step $t$, and $f(\cdot,\cdot):\mathcal{X} \times \mathcal{U} \to \mathcal{X}$ is a continuous function satisfying $f(0,0)=0$. We consider a constrained optimal control problem in which the states and inputs should satisfy time-invariant constraints: $x_t \in X := \{x \in \mathcal{X}| h(x) \leq 0\}$ and  $u_t \in U \subseteq \mathcal{U}$. Here, $h(\cdot): \mathcal{X} \to \mathbb{R}$ is a continuous function that defines the state constraint\footnote{For the constraint defined by multiple inequalities $h_i(x) \leq 0,\; i=1,2,...,I$, we can let $h(x) = \max_{i \in \{1,...,I\}} h_i(x)$. The set $\{x| h(x) \leq 0\}$ is then identical to $\{x| h_i(x) \leq 0,\;i=1,2,...,I\}$, and $h$ will be continuous if each $h_i$ is continuous.}. In most parts of the paper, we assume that $f$ is fully known. However, in Section V.C, we briefly discuss the applicability of our method when there is system uncertainty. Besides, we assume that $X$ is compact and that $U$ is a polytope.  

\textcolor{blue}{The control objective is to regulate a predefined control policy, which could be a policy learned through various methods such as RL \cite{li2023reinforcement}, learning-based MPC \cite{karg2020efficient}, or any other suitable control technique that needs safety guarantees.} \textcolor{blue}{The primary concern is ensuring constraint satisfaction \emph{after} learning, rather than safe exploration \cite{cohen2021safe}, which is not the focus here.}

\subsection{Control barrier functions}
To achieve the objective, we need to design a control policy that can ensure constraint satisfaction at all time steps. A set of states for which such a policy exists, needs to be defined. Usually, this class of sets is called controlled-invariant sets or safe sets. For high-dimensional systems, however, some controlled-invariant sets could have complex representations, making the controller synthesis difficult. A CBF uses the sub-level set of a scalar function to conveniently define the safe set.

\begin{definition}[Control barrier function]\label{CBF definition}
	A continuous function $B(\cdot):\mathcal{X} \to \mathbb{R}$ is called a control barrier function (CBF) with a corresponding safe set $\mathcal{S}_B:=\left\{x \in \mathbb{R}^{n_x}| B(x) \leq 0\right\} \subseteq \mathcal{X}$, if $\mathcal{S}_B$ is non-empty, 
	\begin{equation}\label{CBF0}
		h(x)\leq 0,\;\forall x \in \mathcal{S}_B,
	\end{equation}
	and
	\begin{equation}\label{CBF}
		\textcolor{blue}{\forall x \in \mathcal{S}_B,\;\exists u\in U \text{ s.t. } B(f(x, u)) \leq 0.}
	\end{equation}
	\textcolor{blue}{Furthermore, a CBF is called an exponential CBF if $h(x)\leq B(x),\;\forall x \in \mathcal{S}_B$ and if there exists} a $\beta \in (0,1)$ such that
	\begin{equation}\label{CBFex}
		\textcolor{blue}{\forall x \in \mathcal{S}_B,\;\exists u\in U \text{ s.t. }  B(f(x, u)) \leq \beta B(x).}
	\end{equation}
\end{definition}

Condition \eqref{CBF0} is equivalent to $\mathcal{S}_B \subseteq X$, which is usually assumed in literature \cite{agrawal2017discrete}. With a CBF available, one can generate a safe control policy in $\mathcal{S}_B$ by using the following optimization-based approach:
\begin{align}\label{filter}
	\pi_{\mathrm{safe}}(x)\!=\!&\arg\min_{u\in U}\; ||u-\pi_0(x)||_2\\
	&\mathrm{s.t.}\; B(f(x, u)) \leq 0,  \textcolor{blue}{\mathrm{if}\; B \text{ is a CBF (not exponential)}}\nonumber\\
	&\mathrm{or}\; B(f(x, u)) \leq\beta B(x), \textcolor{blue}{ \mathrm{if}\; B  \text{ is an exponential CBF},}\nonumber
\end{align}
which serves as a safety filter \cite{li2022robust} for any unsafe policy $\pi_0$. The definition of a safe policy is formally given as follows.
\begin{definition}\label{safepolicy}
	A policy $\pi(\cdot): \mathcal{X} \to \mathcal{U}$ is safe in $\mathcal{S}\subseteq  X$ for system \eqref{system} under the state and input constraints $x \in X,\;u \in U$, if $\pi(x) \in U,\;\forall x \in \mathcal{S}$, and for any initial state in $\mathcal{S}$, the state trajectory of \eqref{system} steered by $\pi$ will always stay in $\mathcal{S}$.
\end{definition}

\textcolor{blue}{For general nonlinear systems with state and input constraints, synthesizing a non-conservative CBF is a difficult task. Hand-crafted or application-specific heuristics are mostly used in literature to design candidate CBFs, which can be either unsafe or overly conservative (see \cite[Figure 1]{tonkens2022refining} for an illustrative example).} To deal with this issue, using advanced function approximators to learn a CBF certificate has received much attention (see \cite{dawson2023safe} for a comprehensive survey).

\subsection{Limitations of existing CBFs}

\emph{Problem P1: high online computational complexity.} One main limitation of \eqref{filter} is that it needs to solve a usually non-convex optimization problem in real time. If a CBF is represented by a complex function approximator $B_\theta(\cdot)$ with the parameter $\theta$, solving \eqref{filter} may take much online computation time and result in very sub-optimal solutions. Even if a convex $B_\theta$ is formed, the constraint in \eqref{filter} could be non-convex due to the nonlinearity of $f$. 
\\
\emph{Problem P2: effects of approximation errors.} Another limitation of \eqref{filter} is that the approximation error of $B(\cdot)$ may affect the safety of the system controlled by the optimizer of \eqref{filter} with $B(\cdot)$ replaced by $B_\theta(\cdot)$. Besides, the recursive feasibility of \eqref{filter} is also not guaranteed with $B_\theta(\cdot)$.

\section{State-action control barrier function}
To deal with Problem P1, motivated by Q-learning in RL \cite{bertsekas2019reinforcement}, we propose a novel safety filter in the following form:
\begin{align}\label{filter2}
	\pi_{\mathrm{safe}}(x) = \arg\min \{||u-\pi_0(x)||_2,\;\mathrm{s.t.}\; u\in U,\; Q(x,u) \leq 0\},
\end{align}
where $Q(\cdot,\cdot) : \mathcal{X} \times U \to \mathbb{R}$ is a function of states and actions. We will analyze how to enforce the safety of $\pi_{\mathrm{safe}}$ by imposing conditions on $Q$. To achieve this, we introduce the definition of SACBFs as follows.

\begin{definition}[State-action control barrier function]\label{SACBF definition}
	\textcolor{blue}{A continuous function $Q(\cdot,\cdot) : \mathcal{X} \times U \to \mathbb{R}$ is called a state-action control barrier function (SACBF) with a corresponding safe set $\mathcal{S}_Q$, if the pair $(Q, \; \mathcal{S}_Q)$ satisfies}
	
	\noindent\textcolor{blue}{(i) $\mathcal{S}_Q$ is non-empty, and $h(x) \leq 0,\; \forall x \in \mathcal{S}_Q$,}
	
	\noindent\textcolor{blue}{(ii) $\forall x\in \mathcal{S}_Q,\;\exists u\in U \text{ s.t. } Q(x,u) \leq 0$, and}
	
	\noindent(iii) for any $x \in \mathcal{S}_Q$, any $u \in U$ that satisfies $Q(x,u) \leq 0$ will make $f(x,u) \in \mathcal{S}_Q$.
\end{definition}

\textcolor{blue}{Unlike the definition of standard CBFs, we do not prescribe the form of $\mathcal{S}_Q$ based solely on $Q$. Besides, condition (i) imposes $\mathcal{S}_Q \subseteq X$. The following lemma builds the connection between standard CBFs and SACBFs, and provides an explicit form of $\mathcal{S}_Q$ when the SACBF is derived from a standard CBF.}
\begin{lemma}\label{lemma2}
	(i) For the safety filter \eqref{filter2}, any SACBF $Q$ will render $\pi_{\mathrm{safe}}$ safe in $\mathcal{S}_Q$. In other words, $\mathcal{S}_Q$ is control-invariant.
	
	\textcolor{blue}{(ii) If $B$ is a CBF, $Q(\cdot,\cdot) = B(f(\cdot,\cdot))$ will be an SACBF with the safe set $ \mathcal{S}_B$, i.e., $ \mathcal{S}_Q =  \mathcal{S}_B$.}
\end{lemma}

The proof of these properties is given in Appendix \ref{appendix1}.

%\begin{equation}\label{CBF relation}
%	Q(x,u) = B(f(x,u)),\; \mathrm{or}\; 	Q(x,u) = B(f(x,u))-\beta B(x). 
%\end{equation}

Similarly to designing CBF, the challenge of using \eqref{filter2} is that the explicit form of an SACBF cannot be directly obtained based on Definition \ref{SACBF definition}. We will present a learning-based approach to synthesize SACBFs in Section V.

The advantage of directly approximating $Q$ over approximating $B$ is that we can design a specific structure for $Q_\theta$ to simplify the constraint in \eqref{filter2}, so that the online computational cost of solving \eqref{filter2} is reduced. To achieve this, we can specify the following parameterization: 
\begin{equation}\label{parameterization}
	Q_\theta(x,u) = q_{1,\theta} (x)+ q_{2,\theta} (x) u + u^T Q_{3,\theta} (x) u,\;\mathrm{with}\;Q_{3,\theta} \succeq 0,
\end{equation}
which will make problem \eqref{filter2} a convex QP with linear and convex quadratic constraints. In \eqref{parameterization}, $q_{1,\theta} (\cdot):\mathbb{R}^{n_x} \to \mathbb{R}$, \textcolor{blue}{$q_{2,\theta}(\cdot) :\mathbb{R}^{n_x} \to \mathbb{R}^{n_u}$ and $Q_{3,\theta}(\cdot) :\mathbb{R}^{n_x} \to \mathbb{R}^{n_u \times n_u}$} are parameterized functions with all parameters condensed in $\theta$.

\textcolor{blue}{If the system (1) has a control-affine form, i.e., $f(x,u) = f_x (x) + f_u(x)u$, and the linearized system of (1) around the origin is stabilizable, there always exists an SACBF in the form of \eqref{parameterization}. This is because a quadratic CBF always exists for (1) according to Appendix D, and $Q(\cdot, \cdot)=B(f(\cdot, \cdot))$ is an SACBF if $B$ is a CBF. Additionally, we should note that for general non-affine control systems, a quadratic SACBF is not guaranteed to exist.}

\textcolor{blue}{Our proposed concept of SACBFs is inspired by the definition of state-action (Q) value functions in RL \cite{li2023reinforcement}. Both kinds of functions not only characterize the energy of each state but also quantify the quality of taking each action in each state. In Section V, we will demonstrate that the Q value function for a specific finite-horizon optimal control problem is, in fact, an SACBF.}

\section{Constraint tightening and contractive-set CBFs}

To cope with Problem P2, we propose a method to compute conservative CBFs based on state constraint tightening. In particular, we consider two kinds of conservative CBFs. The first kind includes the CBFs of system \eqref{system} under the tightened state constraint $X_\lambda := \{x \in \mathcal{X}| h(x) + \lambda \leq 0\}$ where $\lambda$ is a positive constraint backoff. In this situation, the CBFs still satisfy \eqref{CBF} or \eqref{CBFex}, while condition \eqref{CBF0} is tightened to $
h(x) +\lambda \leq B(x), \forall x \in \mathcal{S}_B$.

The second one, called the contractive-set CBF, is defined as follows
\begin{definition}[Contractive-set CBF]\label{contractive-set CBF definition}
	A CBF $B(\cdot):\mathcal{X} \to \mathbb{R}$ is called a $\lambda$-contractive-set CBF\footnote{Note that in fact the definition of contractive-set CBFs given here is not consistent with the common definition of contraction mappings \cite{bertsekas2019reinforcement}. Formally speaking, the CBF we define should be called a CBF with a contractive safe set \cite{alessio2007squaring}. However, for compactness we adopt “contractive-set CBF” in the paper.} if there exists a $\lambda>0$ such that
	\begin{equation}\label{contractive}
		\min _{u \in U} B(f(x, u)) \leq -\lambda,\;\forall x \in \mathcal{S}_B.
	\end{equation}
	Similarly, an exponential CBF $B(\cdot):\mathcal{X} \to \mathbb{R}$ is called a $\lambda$-contractive-set exponential CBF if there exist $\lambda>0$ and $\beta \in (0,1)$ such that
	\begin{equation}\label{contractive_ex}
		\min _{u \in U} B(f(x, u)) \leq \beta B(x)-\lambda,\;\forall x \in \mathcal{S}_B.
	\end{equation}
\end{definition}

The definition of contractive-set CBFs is introduced to endow the safe set $\mathcal{S}_B$ with a contractive property. It enforces an energy decay when the state is on the boundary of the safe set. In particular, if $B(\cdot)$ is a $\lambda$-contractive-set CBF, for any $x \in \partial \mathcal{S}_B$, there exists an input $u\in U$ such that $B(f(x,u)) \leq -\lambda$, which means that $x^+ =f(x,u)$ is in the interior of $\mathcal{S}_B$. Such a property can guarantee that any approximation of $B(\cdot)$ is still a valid CBF if the approximation error is sufficiently small.

\section{Approximating SACBFs}

In this section, we propose a method for approximating the SACBF. Inspired by HJ reachability \cite{choi2021robust} and predictive CBF \cite{didier2023approximate}, we propose a comprehensive framework that uses the optimal value functions of a sequence of optimization problems to implicitly represent the CBF sequence $\{B_k\}^{\infty}_{k=1}$. By tuning some parameters, this framework can compute standard CBFs, exponential CBFs, as well as contractive-set CBFs. \textcolor{blue}{Although the exact expression of each $B_k$ is in general intractable to compute, we can use learning-based approaches such as supervised learning and RL \cite{li2023reinforcement} and then use the results of Lemma 1 to approximate SACBFs.}  

\subsection{CBF generator based on reachability analysis}

We consider each $B_k$ as the value function of the following optimization problem: 
\begin{align}\label{CBF generator}
	B_k(x) := &\min _{ \{x_t,u_t\}_{t=0}^{k}} \max\left\{\max _{t \in \mathcal{I}(k-1)} \alpha^t (h(x_t )+\lambda_t), \alpha^k B_0(x_k )\right\} \nonumber\\
	&\quad \mathrm{s.t.}\; x_{t+1}=f\left(x_t, u_t\right),\;\;u_t \in U,\;t \in \mathcal{I}(k-1),\nonumber\\
	&\quad \quad \quad x_0 = x.
\end{align}
In \eqref{CBF generator}, $\mathcal{I}(k-1)=\{0,1,...,k-1\}$, $\alpha \geq 1$, and $B_0$ is a CBF, which, however, could have a very small safe set. \textcolor{blue}{We take $\alpha=1$ if the CBF $B_0$ is not an exponential CBF}, or $\alpha = 1/\beta$ if $B_0$ is an exponential CBF. To solve \eqref{CBF generator}, we need to know the explicit formulation of $B_0$. An LMI-based method that can compute a local quadratic $B_0$ for the nonlinear system \eqref{system} is reported in \cite{wabersich2022predictive}. In Appendix \ref{computingB0}, this method is extended to our situations where additional conditions such as \eqref{contractive}, \eqref{contractive_ex}, and \eqref{strenghten}-\eqref{strenghten2} in the subsequent Theorem \ref{theorem1} are required. \textcolor{blue}{The LMI method to obtain $B_0$ relies on the linearization of both the system and the constraints. The resulting $B_0$ is in general very conservative. In contrast, by introducing the reachability problem \eqref{CBF generator}, we get less conservative CBFs $B_k$, as stated in Theorem \ref{theorem1} below.}

In \eqref{CBF generator}, $\lambda_t\geq 0,\;t \in \mathcal{I}(k-1)$ are tuning parameters. We consider the following three options for choosing $\lambda_t$:

\emph{Option 1:} $\lambda_t = 0$. \emph{Option 2:} $\lambda_t = \lambda$, where $\lambda>0$ is a constant. \emph{Option 3:} $\lambda_t = t\lambda$, where $\lambda>0$ is a constant.

Option 1 means that we are constructing CBFs for the system \eqref{system} under the original state constraints. If Option 2 is chosen, it is seen from \eqref{CBF generator} that we increase the function $h$ to $h+\lambda$, i.e., we tighten the original state constraints to $x \in X_\lambda=\{x \in \mathcal{X}| h(x) + \lambda \leq 0\}$. More conservatively, in Option 3, we require a linear decrease rate for $h$ w.r.t. the time step to construct contractive-set CBFs. 

In the original HJ reachability analysis \cite{choi2021robust}, $B_0(\cdot)$ is chosen as $h(\cdot)$, which is not a CBF. It has been proven in \cite{choi2021robust} that only when $k=\infty$, $B_\infty$ is a CBF. Although this result shows a strong connection between CBFs and HJ reachability, it cannot be used in practice since we cannot solve \eqref{CBF generator} with $k=\infty$. In our case, we require the initial function $B_0$ to be a CBF. As a result, any $B_k,\;k=1,2,...$ will become a CBF, with a non-shrinking safe set as $k$ increases. The following theorem formally states this property, and the proof is given in Appendix \ref{proof of Theorem 1}.

\begin{theorem}\label{theorem1}
	Consider $B_k$ from \eqref{CBF generator}. Suppose that $B_0$ is a (exponential) CBF for \eqref{system} with the state constraint $x\in X$.
	\begin{enumerate}
		\item Let $\lambda_t = 0,\;t \in \mathcal{I}(k-1)$. Then $B_k,\;k=1,2,...$ is a (exponential) CBF for system \eqref{system} with the state constraint $x\in X$.
		
		\item Let $\lambda_t = \lambda>0,\;t \in \mathcal{I}(k-1)$. If $\lambda,B_0$ satisfy 
		\begin{equation}\label{strenghten}
			h(x) +\lambda \leq B_0(x), \forall x \in \mathcal{S}_{B_0},
		\end{equation}
		then $B_k,\;k=1,2,...$ is a (exponential) CBF for system \eqref{system} with the tightened state constraint $x\in X_\lambda=\{x| h(x) + \lambda \leq 0\}$.
		
		\item Let $\lambda_t = t \lambda,\;t \in \mathcal{I}(k-1)$, $\lambda>0$. If $B_0$ is a $\lambda$-contractive (exponential) CBF and $k,\lambda,B_0$ satisfy 
		\begin{equation}\label{strenghten2}
			h(x) +k\lambda \leq B_0(x), \forall x \in \mathcal{S}_{B_0},
		\end{equation}
		then $B_k,\;k=1,2,...$ is a $\lambda$-contractive (exponential) CBF for system \eqref{system} with the state constraint $x\in X$.

		\item In the statements (i)-(iii), $B_k(\cdot)$ is a continuous function in $\mathcal{X}$, and the safe sets satisfy $\mathcal{S}_{B_0} \subseteq \mathcal{S}_{B_1}\subseteq...\subseteq \mathcal{S}_{B_\infty}$. Furthermore, if $f$, $h$, and $B_0$ are Lipschitz continuous, $B_k$ is Lipschitz continuous.
	\end{enumerate}
\end{theorem}

If $B_k$ and $f$ are (Lipschitz) continuous in their domains, then the function $Q$ defined by
\begin{equation}\label{CBF relation}
	Q(x,u) = B_k(f(x,u))
\end{equation}
is also (Lipschitz) continuous in $\mathcal{X} \times U$.

The condition \eqref{strenghten2} indicates that the horizon $k$ should not be selected too large when using \eqref{CBF generator} to generate a contractive-set CBF.

The structure of \eqref{CBF generator} is similar to that in \cite{didier2023approximate,korda2020computing}. The main difference is that we use the maximum of $h(\cdot)+\lambda_t $ and $B_0(\cdot)$ over state trajectories in a finite horizon, while \cite{didier2023approximate,korda2020computing} uses the summation of $\max\{h(\cdot),0\}$ and $\max\{B_0(\cdot),0\}$ over state trajectories in a finite horizon. This difference means that the safe set $\mathcal{S}_{B_k}$ is the zero-sublevel set of $B_k$ in our situation, while the safe set in \cite{didier2023approximate,korda2020computing} is the zero-level set of $B_k$, i.e., the set $\{x\in \mathbb{R}^{n_x}| B_k(x) =0\}$. Besides, in  \cite{didier2023approximate,korda2020computing} the weight on the terminal CBF $B_0$ needs to be carefully selected.

\subsection{Learning SACBFs from CBF samples}\label{learning CBF}
\textcolor{blue}{We assume that the prediction model in the reachability problem \eqref{CBF generator} is known. To approximate SACBFs, we opt for supervised learning, which is more computationally efficient than RL because it can solve \eqref{CBF generator} exactly and efficiently for each state sample.}

After the CBF generator is provided with a fixed $k$, a regression model can be trained to obtain an approximation of an SACBF. 

First, state and input samples are collected in a compact region $\Omega \times U$ of interest. The region $\Omega$ is task-specified and should be contained in the space where the system is physically realistic. If the state space $\mathcal{X}$ of the system is compact and small, letting $\Omega = \mathcal{X}$ is an ideal choice. If $\mathcal{X}$ is unbounded or too large, one possible choice for $\Omega$ is $\Omega = \{x\in\mathcal{X}| B_k(x) \leq \bar{B}\}$ \cite{didier2023approximate}, where $\bar{B}>0$ is a positive constant. In the case of $\alpha >1$, the values of $B_k$ and $Q$ are likely to grow exponentially as $k$ increases, so it is necessary to specify $\Omega$ as $ \{x\in\mathcal{X}| B_k(x) \leq \bar{B}\}$ to avoid approximating probably unbounded $Q$. Besides, various sampling methods \cite{mesbah2022fusion}, such as (quasi) random sampling and sampling from a uniform grid, can be applied. 

After $N$ samples $\{x^{(\mathrm{s})}_i,u^{(\mathrm{s})}_i\}^{N}_{i=1}$ are collected, problem \eqref{CBF generator} is solved with $x$ specified as each $f(x^{(\mathrm{s})}_i,u^{(\mathrm{s})}_i)$. As a result,  $N$ data tuples $\{(x^{(\mathrm{s})}_i,\;u^{(\mathrm{s})}_i,\;Q(x^{(\mathrm{s})}_i,u^{(\mathrm{s})}_i))\}^{N}_{i=1}$ are obtained after substituting $x =x^{(\mathrm{s})}_i,\;u=u^{(\mathrm{s})}_i$ into \eqref{CBF relation}. For general nonlinear systems, \eqref{CBF generator} is usually a nonlinear non-convex optimization problem, which requires a multi-start strategy \cite{rinnooy1987stochastic} to find a sufficiently good optimum. For linear systems with linear or ellipsoidal constraints, \eqref{CBF generator} is a convex quadratically constrained quadratic program (QCQP) and the global optimum can be conveniently obtained via gradient-based optimization methods. For piecewise affine systems with piecewise affine constraints, \eqref{CBF generator} can be regarded as a mixed-integer QCQP, which can be globally solved by the branch-and-bound approach \cite{wolsey1999integer}.

In this work, we use NNs to approximate the SACBF. For the parameterization \eqref{parameterization}, we use three NNs to represent  $ q_{1,\theta},\; q_{2,\theta}$, and $Q_{3,\theta}$ in \eqref{parameterization}, respectively. There are several ways to guarantee the positive-semidefiniteness of $Q_{3,\theta}$. For example, we can further parameterize $Q_{3,\theta}$ by $Q_{3,\theta} = L_{\theta} L^T_{\theta}$, where $L_{\theta} \in \mathbb{R}^{n_u\times n_u} $ is a lower triangular matrix with non-negative diagonal entries. Alternatively, we can parameterize $Q_{3,\theta}$ by $Q_{3,\theta} = P_0 \mathrm{diag}(r_{\theta}) \mathrm{diag}(r_{\theta}) P^T_0$, where $r_{\theta} \in \mathbb{R}^{n_u}$ is the output of an NN and $P_0 \in \mathbb{R}^{n_u\times n_u}$ is a given matrix.

\textcolor{blue}{Finally, $\theta$ is optimized to minimize the mean square
	error between $B_k(f)$ and $Q_\theta$ over all state-action samples, using classical NN training algorithms such as stochastic gradient descent \cite{goodfellow2016deep}.}

\subsection{Performance analysis}

After the approximation $Q_\theta$ has been obtained, it can be integrated into the safety filter \eqref{filter2}, i.e., $Q$ in \eqref{filter2} is replaced by $Q_\theta$. In general, the safety of the policy generated from \eqref{filter2} will not be ensured due to the approximation error. However, we will now demonstrate that we can guarantee safety in the presence of a small approximation error. To achieve this, we need an assumption on the boundedness of this error.
\begin{assumption}\label{A1}
	The approximation error is uniformly bounded in $\Omega \times U$. In other words, there exists a non-negative constant $ \Delta < \infty$ such that
	\begin{equation}\label{key}
		|Q_\theta(x,u)-Q(x,u)| \leq \Delta,\;\forall (x,u) \in \Omega \times U,
	\end{equation}
	where $\Omega$ is a compact set and satisfies $\mathcal{S}_Q \subseteq \Omega$.
\end{assumption}

Assumption \ref{A1} is common in the literature studying performance guarantees of learning-based control \cite{hertneck2018learning,didier2023approximate,he2022approximate}. If the chosen function approximator represents a continuous function, since $Q$ is also continuous, the approximation error is always upper bounded in any compact region. To obtain the upper bound, we can first get the error bound on a finite number of samples, and then extract a statistical estimate \cite{hertneck2018learning} or compute a deterministic bound in the whole region by using the Lipschitz property of $Q$ and $Q_\theta$ \cite{he2022approximate}.

With Assumption \ref{A1}, we modify the safety filter \eqref{filter2} to 
\textcolor{blue}{ \begin{align}\label{filter3}
		\pi_{\theta}(x) =\arg \min_{u\in U} &||u-\pi_0(x)||_2\nonumber\\
		\mathrm{s.t.}\; &Q_\theta (x,u) \leq 0 \text{ for Options 1 and 2 }\nonumber\\
		&Q_\theta (x,u) \leq -\lambda+\Delta \text{ for Option 3}.
\end{align}}
The following theorem, the proof of which is given in Appendix \ref{proof of Theorem2}, characterizes the safety performance of the policy $\pi_{\theta}$.
\begin{theorem}[Safety and recursive feasibility]\label{theorem2}
	Consider the system \eqref{system} controlled by $\pi_{\theta}$, the SACBF $Q$ from \eqref{CBF relation}, and the safety filter \eqref{filter3}. Suppose that Assumption \ref{A1} holds.
	
	(i) If $B_k$ is a CBF for the original state constraint, for any initial state $x_0 \in  \mathcal{S}_{B_k}$ that makes problem \eqref{filter3} recursively feasible,  the closed-loop system $x_{t+1}=f\left(x_t, \pi_{\theta}(x_t)\right), \; t=0,1, \ldots$ has the maximum constraint violation $\Delta$, i.e., $\max_{t \in \mathcal{I}(\infty)}\{h(x_t)\} \leq \Delta$.
	
	(ii) If $B_k$ is a CBF for the tightened state constraint $x \in X_\lambda$ and $\Delta \leq \lambda$, for any initial state $x_0 \in  \mathcal{S}_{B_k}$ that makes problem \eqref{filter3} recursively feasible,  the closed-loop system $x_{t+1}=f\left(x_t, \pi_{\theta}(x_t)\right)$ always satisfies the state constraint $x\in X$, i.e., $\max_{t \in \mathcal{I}(\infty)}\{h(x_t)\} \leq 0$.
	
	(iii) If $B_k$ is a $\lambda$-contractive-set CBF for the original state constraint and $\Delta \leq \lambda/2$, problem \eqref{filter3} will be recursively feasible for all initial states in $\mathcal{S}_{B_k}$ and the closed-loop system $x_{t+1}=f\left(x_t, \pi_{\theta}(x_t)\right)$ always satisfies the state constraint $x\in X$, i.e., $\max_{t \in \mathcal{I}(\infty)}\{h(x_t)\} \leq 0$.
\end{theorem}

In other words, in Option 2 of \eqref{CBF generator}, we get a CBF $B_k$ for the tightened constraint $x \in X_\lambda$. The policy $\pi_{\theta}$ will always make the system satisfy the original state constraint if \eqref{filter3} is always feasible. However, the feasibility of problem \eqref{filter3} is only guaranteed for the initial state, i.e., problem \eqref{filter3} can be infeasible for some subsequent states. Practically, ones need to perform offline some statistical or deterministic verification \cite{hertneck2018learning} to analyze the recursive feasibility of \eqref{filter3}. Otherwise, as recursive feasibility of \eqref{filter3} cannot be predicted in advance, one needs to do some relaxation for the constraint in \eqref{filter3} once infeasibility is observed. In comparison, in Option 3 of \eqref{CBF generator}, we get a $\lambda$-contractive-set CBF $B_k$. An induced feature is the recursive feasibility of problem \eqref{filter3}. As a result, under the sufficiently small approximation error, the SACBF approximation $Q_\theta$ will render $\pi_{\theta}$ a safe policy according to Definition \ref{safepolicy}.

\textcolor{blue}{The proposed constraint tightening approach naturally has robustness to model uncertainty. In particular, suppose that (1) is a nominal system, and the real system is $x_{t+1} = f_\mathrm{r}(x_t,u_t)$. The mismatch between $f$ and $f_\mathrm{r}$ can be handled similarly to the learning error of $Q$. Formally, we have the following corollary:}

\textcolor{blue}{\begin{corollary}[Robustness to model uncertainty]
		Suppose that Assumption \ref{A1} is fulfilled and the nominal system satisfies $\left|f(x, u)-f_\mathrm{r}(x, u)\right| \leq \kappa, \forall(x, u) \in \Omega \times U $, where $\kappa \geq 0$. Then, if $B_k$ is a $\lambda$-contractive-set CBF for the nominal system under the original state constraint and $\lambda \geq 2 \Delta + \mathrm{L}_B \kappa$ with $\mathrm{L}_B$ the Lipschitz constant of $B_k$, problem (21) will be recursively feasible for the real system with all initial states in $\mathcal{S}_{B_k}$ and the closed-loop system $x_{t+1}=f_\mathrm{r}\left(x_t, \pi_{\theta}(x_t)\right)$ always satisfies the state constraint $x\in X$, i.e., $\max_{t \in \mathcal{I}(\infty)}\{h(x_t)\} \leq 0$.
\end{corollary}	}

\textcolor{blue}{The proof is similar to the proof of (iii) of Theorem \ref{theorem2} and is thus omitted here.}

\subsection{Discussion on SACBFs}
\textcolor{blue}{In addition to the computational benefit bought by introducing SACBFs, another advantage is that it enables the synthesis of both safe controllers and safe certificates even when the explicit form of the model is unknown. The similarity between SACBFs and Q value functions in RL indicates that it is possible to use RL methods \cite{fisac2019bridging} such as Q value iteration or Q-learning to learn SACBFs directly without knowing the explicit model. It should be noticed that even though there are several data-driven methods to synthesize standard CBFs from black-box models \cite{robey2020learning,bonzanini2022scalable}, a prediction model is necessary to implement the safety filter (5). In comparison, the proposed SACBF provides the possibility of bridging the gap between model-free learning control and CBF-based safety filters. This technological roadmap, which we refer to as direct data-driven safe control and plan to explore in future work, has been recognized as a promising direction in recent survey papers \cite{dawson2023safe,wabersich2023data}.}

One limitation of the parameterization \eqref{parameterization} is that it sacrifices the universal approximation property \cite{hanin2019universal} of NNs, since it restricts $Q_\theta$ to be quadratic on $u$. This may result in a large error bound $\Delta$ and cause constraint violations according to Theorem \ref{theorem2}. As our main motivation for introducing state-action CBFs is to improve online computational efficiency, such a sacrifice is acceptable. Conversely, fully parameterizing $Q_\theta$ with a continuous NN, possessing universal approximation capabilities, increases the online computational complexity and introduces challenges in finding the global optimum of \eqref{filter3}. In conclusion, when applying an approximation to the SACBF, there is typically a natural trade-off between online computational efficiency and safety. 

\textcolor{blue}{Note that a parametrization of an SACBF $Q$, instead of parameterizing a standard CBF $B$, may increase the approximation error on $B(f)$, especially when $Q$ is parameterized in a quadratic form. This error can result in the learned $Q$ failing to satisfy the conditions of a valid SACBF. To address this issue, we have proposed the constraint tightening approach (Section IV) to ensure that the learning-based approach can still generate a valid SACBF under mild approximation errors. However, a quadratic parametrization of $Q$ typically requires more stringent constraint tightening, which can lead to increased conservatism in the learned barrier certificates. }

\textcolor{blue}{The conservatism of the learned SACBF $Q_\theta$ as well as the policy $\pi_{\theta}$ is strongly influenced by $\Delta$. According to Theorem \ref{theorem2}, to guarantee the recursive feasibility of \eqref{filter3} and the safety of $\pi_{\theta}$, a larger $\Delta$ requires a larger $\lambda$ in \eqref{CBF generator}. This, in turn, increases the conservatism of the target $Q$. In practice, to minimize the conservatism, one can start with a small $\lambda$ and iteratively increase it until Assumption \ref{A1} and $\Delta \leq \lambda / 2$ are satisfied. Increasing the approximation capability of the chosen function approximator, as discussed in \cite{barron1994approximation}, can also be beneficial.}

\textcolor{blue}{We focus on discrete-time nonlinear systems because our work targets safety in learning-based control methods, like RL and supervised learning of MPC policies, which typically use discrete-time models \cite{rawlings2017model,li2023reinforcement}. Additionally, safety filters for discrete-time systems are computationally more challenging, leading to nonlinear programs, whereas continuous-time systems involve simpler convex quadratic programs \cite{agrawal2017discrete}. Therefore, the computational benefits of introducing SACBFs are more evident in the context of discrete-time nonlinear systems.}

\textcolor{blue}{Compared to learning standard CBFs \cite{robey2020learning}, the sampling complexity of learning the proposed SACBFs increases mildly because the domain of the function to be approximated only changes from $\mathbb{R}^{n_x}$ to $\mathbb{R}^{n_x+n_u}$. More importantly, advanced sampling strategies, such as active learning \cite{bonzanini2022scalable} and counterexample-guided inductive synthesis \cite{abate2022neural}, can be utilized to improve the scalability of approximating $Q$.}

\section{Case study}

\begin{table*}[]
	\scriptsize
	\centering
\caption{Performance of different controllers. }%
\begin{tabular}{|ccccccccccll|}
	\hline
	\multicolumn{1}{|c|}{\multirow{3}{*}{Safety filters}} & \multicolumn{11}{c|}{Basic control policies}                                                                                                                                                        \\ \cline{2-12} 
	\multicolumn{1}{|c|}{}                                & \multicolumn{3}{c|}{Learning MPC}                            & \multicolumn{3}{c|}{ADP}                                         & \multicolumn{5}{c|}{LQR}                                         \\ \cline{2-12} 
	\multicolumn{1}{|c|}{}                                & Safety rate                  & \multicolumn{1}{c|}{CPU time} & \multicolumn{1}{c|}{Cost} & \multicolumn{1}{c}{Safety rate} & \multicolumn{1}{c|}{CPU time} & \multicolumn{1}{c|}{Cost} & \multicolumn{1}{c}{Safety rate} & \multicolumn{1}{c|}{CPU time}& \multicolumn{3}{c|}{Cost}  \\ \hline
	\multicolumn{1}{|c|}{No filter}                       & \multicolumn{1}{c}{63.52\%} & \multicolumn{1}{c|}{0.024 ms} & \multicolumn{1}{c|}{\textcolor{blue}{99.69}}& \multicolumn{1}{c}{36.66\%}     & \multicolumn{1}{c|}{0.024 ms} &\multicolumn{1}{c|}{\textcolor{blue}{125.11}} & \multicolumn{1}{c}{56.99\%}     & \multicolumn{1}{c|}{0.023 ms} & \multicolumn{3}{c|}{\textcolor{blue}{122.10}} \\ \hline
	\multicolumn{1}{|c|}{Standard CBF}                    & \multicolumn{1}{c}{63.52\%} & \multicolumn{1}{c|}{2.3 ms}   &\multicolumn{1}{c|}{\textcolor{blue}{126.18}}& \multicolumn{1}{c}{39.93\%}     & \multicolumn{1}{c|}{2.4 ms}   &\multicolumn{1}{c|}{\textcolor{blue}{161.82}}& \multicolumn{1}{c}{57.17\%}     & \multicolumn{1}{c|}{2.3 ms}  & \multicolumn{3}{c|}{\textcolor{blue}{120.00}} \\ \hline
	\multicolumn{1}{|c|}{Quad. SACBF}               & \multicolumn{1}{c}{70.05\%} & \multicolumn{1}{c|}{1 ms}   &\multicolumn{1}{c|}{\textcolor{blue}{131.76}}  & \multicolumn{1}{c}{47.19\%}     & \multicolumn{1}{c|}{1.1 ms} &\multicolumn{1}{c|}{\textcolor{blue}{161.83}}  & \multicolumn{1}{c}{65.52\%}     & \multicolumn{1}{c|}{0.9 ms} &\multicolumn{3}{c|}{\textcolor{blue}{115.80}}  \\ \hline
	\multicolumn{1}{|c|}{Nonlin. SACBF}               & \multicolumn{1}{c}{65.15\%} & \multicolumn{1}{c|}{2.5 ms}  &\multicolumn{1}{c|}{\textcolor{blue}{132.22}} & \multicolumn{1}{c}{40.19\%}     & \multicolumn{1}{c|}{3.0 ms}  &\multicolumn{1}{c|}{\textcolor{blue}{168.19}} & \multicolumn{1}{c}{60.80\%}     & \multicolumn{1}{c|}{2.9 ms}  &\multicolumn{3}{c|}{\textcolor{blue}{137.30}} \\ \hline
	\multicolumn{1}{|c|}{Quad tightened SACBF}            & \multicolumn{1}{c}{98.73\%} & \multicolumn{1}{c|}{1.0 ms}  &\multicolumn{1}{c|}{\textcolor{blue}{97.29}} & \multicolumn{1}{c}{98.73\%}     & \multicolumn{1}{c|}{1.0 ms}  &\multicolumn{1}{c|}{\textcolor{blue}{99.44}} & \multicolumn{1}{c}{92.92\%}     & \multicolumn{1}{c|}{1.0 ms}&\multicolumn{3}{c|}{\textcolor{blue}{91.47}}   \\ \hline
	\multicolumn{1}{|c|}{Nonlin. tightened SACBF}            & \multicolumn{1}{c}{82.40\%} & \multicolumn{1}{c|}{2.7 ms}  &\multicolumn{1}{c|}{\textcolor{blue}{117.26}} & \multicolumn{1}{c}{82.03\%}     & \multicolumn{1}{c|}{2.7 ms} &\multicolumn{1}{c|}{\textcolor{blue}{116.56}}  & \multicolumn{1}{c}{90.56\%}     & \multicolumn{1}{c|}{3.0 ms} &\multicolumn{3}{c|}{\textcolor{blue}{116.92}}  \\ \hline
	\multicolumn{1}{|c|}{Quad. contr. SACBF}            & \multicolumn{1}{c}{100\%}   & \multicolumn{1}{c|}{0.9 ms}  &\multicolumn{1}{c|}{\textcolor{blue}{98.84}} & \multicolumn{1}{c}{100\%}       & \multicolumn{1}{c|}{1.0 ms}  &\multicolumn{1}{c|}{\textcolor{blue}{101.91}}& \multicolumn{1}{c}{100\%}       & \multicolumn{1}{c|}{0.9 ms} &\multicolumn{3}{c|}{\textcolor{blue}{93.10} } \\ \hline
	\multicolumn{1}{|c|}{Nonlin. contr. SACBF}            & \multicolumn{1}{c}{96.91\%} & \multicolumn{1}{c|}{2.1 ms} &\multicolumn{1}{c|}{\textcolor{blue}{110.44}}  & \multicolumn{1}{c}{92.20\%}     & \multicolumn{1}{c|}{2.0 ms} &\multicolumn{1}{c|}{\textcolor{blue}{128.99}}  & \multicolumn{1}{c}{94.57\%}     & \multicolumn{1}{c|}{3.0 ms} &\multicolumn{3}{c|}{\textcolor{blue}{109.97}}  \\ \hline
	\multicolumn{1}{|c|}{\multirow{2}{*}{MPC}}            & \multicolumn{3}{c|}{Horizon=5}                               & \multicolumn{3}{c|}{Horizon=6}                                   & \multicolumn{5}{c|}{Horizon=7}                                   \\ \cline{2-12} 
	\multicolumn{1}{|c|}{}                                & \multicolumn{1}{c}{70.24\%} & \multicolumn{1}{c|}{38.2 ms} &\multicolumn{1}{c|}{\textcolor{blue}{98.55}}& \multicolumn{1}{c}{95.83\%}     & \multicolumn{1}{c|}{37.6 ms}&\multicolumn{1}{c|}{\textcolor{blue}{97.43}}  & \multicolumn{1}{c}{100\%}       & \multicolumn{1}{c|}{38.5 ms} &\multicolumn{3}{c|}{\textcolor{blue}{96.86}} \\ \hline
\end{tabular}
\end{table*}

We validate the proposed methods for an active inverted pendulum interacting with elastic walls, which is a piecewise affine system \cite{he2023approximate}. The detailed model can be found in Appendix \ref{case study details}. The simulations are conducted in MATLAB R2021a on an AMD Core R7-5800H CPU @3.20GHz machine. All optimization problems involved in the policy filters \eqref{filter} and \eqref{filter3} are solved using the Matlab function “fmincon” and the active-set optimization algorithm. The problem \eqref{CBF generator} and the MPC problem are transformed equivalently into mixed-integer quadratic programming problems and are solved using Gurobi \cite{gurobi2021gurobi}. 

The system contains the state $x = [\theta \; \dot\theta]^T$ where $\theta$ is the pendulum angle, and the input $u$, which is the torque acting on the bottom end of the pendulum. The state and input constraints\footnote{\textcolor{blue}{The theoretical analysis in the previous sections is applicable to nonlinear constraints. However, for this case study, we focus on linear constraints for illustrative purposes.}} are given by $|x_1| \leq 0.15$, $|x_2| \leq 1$, and $|u| \leq 4$. The system is discretized using the explicit Euler method with time step 0.05s.

Using the approach in Appendix \ref{computingB0} and letting $\lambda = 0.2$ for Option 2 and $\lambda = 0.05$ for Option 3, we obtain the initial CBF $B_0(x) = x^T P x-1$ with $P = \left[\begin{array}{rr}
	124.74 & 7.44 \\
	7.44 & 2.24
\end{array}\right]$, $\left[\begin{array}{rr}
	283.40 & 20.25 \\
	20.25 & 4.69
\end{array}\right]$, and $\left[\begin{array}{rr}
	522.39 & 25.69 \\
	25.69 & 5.39
\end{array}\right]$ for Options 1, 2, and 3, respectively. We sample from the state-input space $
\left\{(x, u)\big|\;| x_1|\leq 0.16,| x_2|\leq 1.1,| u | \leq 4\right\}$ on a uniform grid and we obtain $40^3$ samples $\{(x_i^{(\mathrm{s})}, u_i^{(\mathrm{s})})\}^{40^3}_{i=1}$. The successor state $f\left(x_i^{(\mathrm{s})}, u_i^{(\mathrm{s})}\right)$ of each sample $\left(x_i^{(\mathrm{s})}, u_i^{(\mathrm{s})}\right)$ is fed to the CBF generator \eqref{CBF generator} with $k = 7$. Only those samples satisfying $B_k (f(x_i^{(\mathrm{s})}, u_i^{(\mathrm{s})}))\leq 10$ are collected. After this procedure, 42109, 38609 and 37027 samples are collected in each option. 

We compare SACBFs and standard CBFs. We consider 4 different kinds of safety certificates: (i) SACBF for the original state constraint; (ii) SACBF for the tightened state constraint $|x_1| \leq 0.14$, $|x_2| \leq 0.9$; (iii) Contractive-set SACBF; (iv) Standard CBF for the original state constraint. For the learning, we use NNs with three hidden hyperbolic tangent layers containing 16, 64 and 8 neurons. We parameterize each standard CBF by an NN. For the parameterization of each SACBF, we use \eqref{parameterization} or fully parameterize it by an NN. Besides, we also consider 4 different control policies. The first three include (i) supervised learning of the MPC policy \cite{karg2020efficient}, (ii) the approximate dynamic programming (ADP) policy \cite{he2023approximate}, which is a kind of RL policy, and (iii) the LQR policy for the linearized system around the origin. These policies may be unsafe and are thus taken as $\pi_0$ in \eqref{filter} and  \eqref{filter3}. The learning-based control policies (i) and (ii) are also represented by NNs. All NNs are trained via the stochastic gradient descent algorithm \cite{goodfellow2016deep}.  \textcolor{blue}{The main reason for choosing these three particular policies is that they can achieve sub-optimal control objectives using very limited computational resources. However, they generally cannot guarantee constraint satisfaction. Therefore, we use these three policies as the baseline to demonstrate that the proposed safety filter can enhance the rate of constraint satisfaction for these policies.} The fourth control policy is implicit MPC. One can refer to Appendix \ref{case study details} for a detailed description of these policies.

To test all the control methods, we uniformly randomly sample initial states from the state sample set $\{x_i^{(\mathrm{s})}\}^{40^3}_{i=1}$ and select a total of 551 initial states $x$ that satisfy $B_k(x)\in (-0.1,0)$. These initial states are contained in the safe set of $B_k$, and thus can be steered to the origin without constraint violation. Meanwhile, since the values of $B_k$ taken at these states are close to zero, these initial states are near the boundary of the safe set. Therefore, they are in general more difficult to stabilize than any other states in the safe set. For each initial state, we simulate the closed-loop systems for 50 time steps.

The simulation results are shown in Table 1. \textcolor{blue}{The performance metrics include the rate of safety, the average CPU time for computing the control input per time step, and the total cost $\sum_{t=0}^{50} C(x_t,u_t)$ with the stage cost the same as that in the considered basic controllers.} The safety rate is defined as the number of initial states leading to a safe closed-loop trajectory divided by the total number of initial states. \textcolor{blue}{Overall, the table indicates that incorporating appropriate safety filters, especially quadratic and nonlinear SACBFs, can significantly improve the safety performance of control policies without excessively compromising on CPU time and total cost.} \textcolor{blue}{The online computational benefit of using the SACBF with the quadratic parametrization \eqref{parameterization} primarily stems from the fact that solving a convex quadratic program is typically faster than solving a nonlinear program with the same number of constraints and decision variables, especially when the nonlinear constraints involve an NN.} \textcolor{blue}{The maximum performance loss is about 34\%, which occurs when applying the nonlinear SACBF to the ADP controller. Considering that the main objective of our work is to impose safety using low computational resources, such a performance degradation is acceptable.} \textcolor{blue}{The inferior performance of the nonlinear contractive-set SACBF compared to its quadratic counterpart with respect to constraint satisfaction is primarily due to the suboptimal solutions when solving \eqref{filter3}. In other words, the assumption of perfect nonlinear optimization does not hold in practical scenarios, leading to the rare instances of constraint violations observed in Table 1.} It should be noticed that the computational advantage of the quadratic SACBF will be larger if multi-start nonlinear optimization is used to solve \eqref{filter} or \eqref{filter3} with a fully parameterized CBF.

\begin{figure}\label{state trajectories}
	\centering
	\subfloat[State trajectories and safe set for the SACBF.]{\includegraphics[width=120pt,clip]{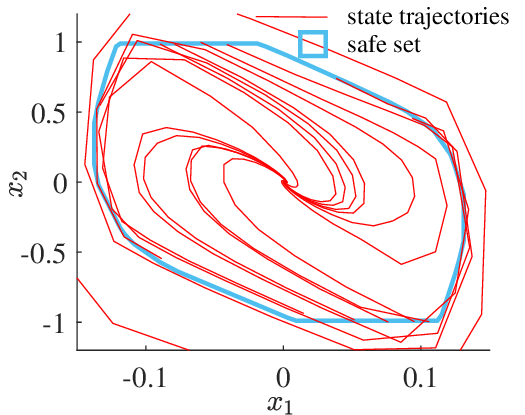}}
	\hfil
	\subfloat[State trajectories and safe set for the contractive-set SACBF.]{\includegraphics[width=120pt,clip]{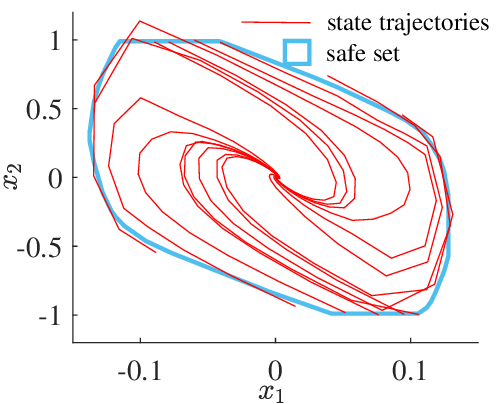}}
	\caption{\textcolor{blue}{Closed-loop trajectories of the learning MPC controller under the safety filter \eqref{filter3}.}}
\end{figure}

\textcolor{blue}{In Figure 1, we demonstrate the state-trajectories of the closed-loop system with the policy $\pi_\theta$ from \eqref{filter3}, with $\pi_0$ being the learning MPC controller and $Q_\theta$ being the nonlinear SACBF (left) or the nonlinear contractive-set SACBF (right). Without constraint tightening, the system evolution records several constraint violations, as evidenced by the 6th line of Table 1. Conversely, the contractiveness property of SACBFs ensures that almost all state trajectories remain within its safe set.}

% \usepackage{multirow}
% \usepackage{multirow}
% \usepackage{multirow}

%In the second case study, we consider a more complex adaptive cruise control problem in which we drive 3 cars (followers) to follow a leading car in a highway environment. The objective is to maintain the distance $d_{i,i+1},\;i=0,1,2,3$ between each too adjacent cars $i,\;i+1$ to 20 m, while satisfying some physical constraints. The system has 6 state variables including the position and velocity of each car, as well as 3 input traction/brake forces acting on each car. The constraints are comprised of the limit on the maximum velocity 35 m/s, the constraint $10 \leq d_{i,i+1}\leq 30,\;i=0,1,2,3$ on the distance between two adjacent vehicles, and the maximum force limit of 3700 N. For the detailed modelling one can see Appendix \ref{case study details}.
%
%Following the design procedure in the first example, we use a policy NN with three hidden hyperbolic tangent layers containing 32, 128 and 16 neurons to approximate MPC. We also learn the previously mentioned four types of safety certificates, represented by NNs with the same structure as the policy NN. In this example, we focus more on the computational advantage brought by the state-action CBF.

\section{Conclusions and future work}
This paper has presented a new type of CBFs, called SACBFs, that can synthesize safe controllers for general constrained nonlinear systems with a very small online computational overhead. We have also developed a new approach to learn the proposed CBF from data, and proposed a constraint tightening approach to improve the robustness of the learned CBFs to learning errors.

In the future, we will first focus on learning the proposed CBFs for large-scale problems in a computationally efficient manner. We will also investigate how the proposed CBFs can be exploited to guide the learning of the basic policy. This will help to reduce the potential degradation of other control performance measure caused by the safety filter. \textcolor{blue}{We will investigate whether undesired equilibria might arise when applying the proposed SACBF-based safety filter, as observed for continuous-time systems.} More importantly, as the proposed safety filter does not need any information about the model, it can be applied in model-free cases provided that the SACBFs can be learned without knowing the model. As model-free (reinforcement) learning control has received much attention recently, the SACBF approach is an important future research direction to achieve safe learning-based control in a completely model-free manner.

\bibliographystyle{IEEEtran}
\bibliography{references}

\appendix
\subsection{Proof of Lemma \ref{lemma2}}\label{appendix1}
\textcolor{blue}{(i) Based on the condition (ii) of Definition \ref{SACBF definition}, for any $x_0 \in \mathcal{S}_Q$, there exists a $u_0 \in U$ such that $Q(x,u) \leq 0$. This means that problem \eqref{filter2} is feasible when $x = x_0$. As $Q(x_0, u_0) \leq 0$ implies $f(x_0,u_0) \in \mathcal{S}_Q$, the control-invariance of $\mathcal{S}_Q$ follows. Consequently, problem \eqref{filter2} is recursively feasible for the initial state $x_0$. Due to the arbitrariness of $x_0$, $\pi_{\mathrm{safe}}$ is safe according to Definition \ref{SACBF definition}.}

\textcolor{blue}{(ii) By specifying $Q(\cdot,\cdot) = B(f(\cdot,\cdot))$ and $\mathcal{S}_Q = \mathcal{S}_B$, we have that $\mathcal{S}_Q$ and $Q$ satisfy the conditions (i) and (ii) of Definition \ref{SACBF definition}. Furthermore, for any $ x_0 \in \mathcal{S}_Q $, consider any $u_0$ such that $Q(x_0,u_0) \leq 0$. We have $B(f(x_0,u_0)) \leq 0$, which further implies $f(x_0,u_0) \in \mathcal{S}_Q$. In addition, as $B$ and $f$ are continuous, $Q$ is also continuous.\hfill$\square$}

\subsection{Proof of Theorem \ref{theorem1}}\label{proof of Theorem 1}
In the following, we will consider the case when $B_0$ is an exponential CBF, i.e., CBF satisfying the additional condition \eqref{CBFex}. The case when $B_0$ is a general CBF satisfying \eqref{CBF} can be analyzed similarly to the case when $B_0$ is an exponential CBF.

The proofs of the first and second statements can be combined, by considering $\lambda_t = \lambda$ and $\lambda \geq 0$. For any $x_0$ such that $B_k(x_0) \leq 0$, letting $(x^*_{t+1}, u^*_t),\;t \in \mathcal{I}(k-1)$ be any one of the optimal solutions to \eqref{CBF generator} when $x = x_0$, we have
\begin{equation}\label{a0}
	h(x^*_t)+\lambda \leq B_k(x_0) \leq 0,\;t=1,2,...,k-1 \;\mathrm{and}\; B_0(x^*_k) \leq 0.
\end{equation}
As $B_0$ is an exponential CBF and $B_0(x^*_k) \leq 0$, we have (i) $h(x^*_k )+\lambda \leq B_0(x^*_k)$ based on \eqref{strenghten}, and (ii) there exists an input $u^*_k \in U$, such that $x^*_{k+1} = f(x^*_k,u^*_k)$ will make $B_0(x^*_{k+1}) \leq \beta B_0(x^*_k)$. Now, we consider the value of $B_k(x^*_1)$. Based on the above discussion, it is clear that the trajectory $(x^*_{t+1}, u^*_t),\;t=1,2,...,k$ is feasible for problem \eqref{CBF generator} when $x = x^*_1$. Due to the optimality of $B_k(x^*_1)$, we have
\begin{align}\label{a1}
	B_k(x^*_1)& \leq \max\left\{\max _{t \in \mathcal{I}(k-1)} \alpha^{t} (h(x^*_{t+1} )+\lambda), \alpha^k B_0(x^*_{k+1} )\right\} \nonumber\\
	&\leq \frac{1}{\alpha} \max\left\{\max _{t \in \mathcal{I}(k)} \alpha^t (h(x^*_t )+\lambda), \alpha^{k+1} B_0(x^*_{k+1} )\right\} \nonumber\\
	&\leq \frac{1}{\alpha} \max\left\{\max _{t \in \mathcal{I}(k-1)} \alpha^t (h(x^*_t )+\lambda), \alpha^{k} B_0(x^*_{k}),  \underbrace{\alpha^{k}(h(x^*_k )+\lambda)}_{\leq \alpha^{k} B_0(x^*_k)} \right\} \nonumber\\
	& \leq \frac{1}{\alpha} \max\left\{B_k(x_0), \alpha^{k} B_0(x^*_k)\right\} \nonumber\\
	& \leq \beta B_k(x_0).
\end{align}
In \eqref{a1}, the second inequality is true because we add the term $h(x^*_0)+\lambda$ to the inner max block. The third inequality follows from $B_0(x^*_{k+1}) \leq \frac{1}{\alpha} B_0(x^*_k)$. The fourth inequality holds because the maximization of the first and second terms in the outer max operation equals $B_k(x_0)$. The last inequality holds because $\alpha^{k} B_0(x^*_k) \leq B_k(x_0)$, according to \eqref{CBF generator}. From \eqref{a1}, we can conclude that $\min _{u \in U} B_k(f(x_0, u)) \leq \beta B_k(x_0)$. This, together with \eqref{a0}, shows that $B_k$ is an exponential CBF for system \eqref{system} under the state constraint $x \in X_\lambda=\{x| h(x) + \lambda \leq 0\}$, since $x_0$ is selected arbitrarily in $\mathcal{S}_{B_k}$. Therefore, by specifying $\lambda = 0$ we prove the first statement, and by letting $\lambda > 0$ we prove the second statement.

Next, we will prove the third statement of Theorem \ref{theorem1}. Similarly to the proof of the first two statements, we consider any $x_0$ such that $B_k(x_0) \leq 0$. Denoting by $(x^*_{t+1}, u^*_t),\;t \in \mathcal{I}(k-1)$ any one of the optimal solutions to \eqref{CBF generator} when $x = x_0$, we have
\begin{equation}\label{a2}
	h(x^*_t)+\lambda_t \leq B_k(x_0) \leq 0,\;t=1,2,...,k-1 \;\mathrm{and}\; B_0(x^*_k) \leq 0.
\end{equation}
Since $B_0$ is a $\lambda$-contractive-set exponential CBF and $B_0(x^*_k) \leq 0$, we have (i) $h(x^*_k )+ k \lambda \leq B_0(x^*_k)$, and (ii) there is a control input $u^*_k \in U$ such that the successor state $x^*_{k+1} = f(x^*_k,u^*_k)$ ensures that $B_0(x^*_{k+1}) \leq \beta B_0(x^*_k)-\lambda$. Then, owning to the optimality of $B_k(x^*_1)$, we have
\begin{align}\label{a3}
	B_k(x^*_1)& \leq \max\left\{\max _{t \in \mathcal{I}(k-1)} \alpha^{t} (h(x^*_{t+1} )+t\lambda), \alpha^k B_0(x^*_{k+1} )\right\} \nonumber\\
	& \leq \frac{1}{\alpha} \max\left\{\max _{t \in \mathcal{I}(k)} \alpha^t (h(x^*_t )+(t-1)\lambda), \alpha^{k+1} B_0(x^*_{k+1} ) \right\} \nonumber\\
	&\leq -\lambda + \frac{1}{\alpha} \max\left\{\max _{t \in \mathcal{I}(k)} \alpha^t (h(x^*_t )+t\lambda), \alpha^{k+1} (B_0(x^*_{k+1} ) + \lambda)\right\} \nonumber\\
	&\leq -\lambda + \frac{1}{\alpha} \max\left\{\max _{t \in \mathcal{I}(k-1)} \alpha^t (h(x^*_t )+t\lambda), \alpha^{k} B_0(x^*_{k} ), \alpha^k (h(x^*_k )+k\lambda) \right\} \nonumber\\
	& \leq -\lambda + \frac{1}{\alpha} \max\left\{B_k(x_0), \alpha^{k} B_0(x^*_k)\right\} \nonumber\\
	& \leq -\lambda + \beta B_k(x_0).
\end{align}
In \eqref{a3}, the second line is true because of the additional introduced item $h(x^*_0)-\lambda$ in the inner max block. The forth inequality holds because of $B_0(x^*_{k+1}) \leq \beta B_0(x^*_k)-\lambda$. The fifth inequality follows from $\max\left\{\max _{t \in \mathcal{I}(k-1)} \alpha^t (h(x^*_t )+t\lambda), \alpha^{k} B_0(x^*_{k} )\right\} = B_k(x_0)$. The last inequality holds because $\alpha^{k} B_0(x^*_k) \leq B_k(x_0)$, according to \eqref{CBF generator}. Together with \eqref{a2}, \eqref{a3} implies that $B_k$ is a $\lambda$-contractive-set exponential CBF for system \eqref{system} with the state constraint $x\in X$.

Finally, we prove the last statement of Theorem \ref{theorem1}. In all the cases of $\lambda_t=0$, $\lambda_t=\lambda$, and $\lambda_t = t\lambda$, we consider the state and input sequences $\{x^*_t\}^{k+1}_{t=0}$, $\{u^*_t\}^{k}_{t=0}$, where $x^*_0 =x_0$. The sequences satisfy 
\begin{equation}\label{a3.5}
	B_k(x^*_0) \leq 0,\; h(x^*_t)+\lambda_t \leq 0,\;t \in \mathcal{I}(k),\; B_0(x^*_{k+1})\leq 0.
\end{equation}
From the last inequality in \eqref{a3.5} and Definitions \ref{CBF definition} and \ref{contractive-set CBF definition}, we know that $h(x^*_{k+1})+\lambda_{k+1} \leq 0$ and that there exists a $u^*_{k+1} \in U$ such that the value of $B_0$ at $x^*_{k+2} = f(x^*_{k+1},u^*_{k+1})$ is smaller than or equal to zero. Meanwhile, note that the sequences $\{x^*_t\}^{k+2}_{t=0}$ and $\{u^*_t\}^{k+1}_{t=0}$ constitute a feasible solution to problem \eqref{CBF generator} with $k+1$, so we get an upper bound on $B_{k+1}(x_0)$ as $B_{k+1}(x_0)\leq 0$. Due to the arbitrariness of $x_0$ in $\mathcal{S}_{B_k}$, we can conclude that $\mathcal{S}_{B_k} \subseteq \mathcal{S}_{B_{k+1}}$, which further by recursion proves that $\mathcal{S}_{B_0} \subseteq \mathcal{S}_{B_1}\subseteq...\subseteq \mathcal{S}_{B_\infty}$.

For the continuity of the value function, we use a similar analysis structure as in \cite{korda2020computing}. For any $x_0, y_0 \in \mathcal{X}$, let $(x^*_{t+1}, u^{x*}_t),(y^{*}_{t+1}, u^{y*}_t),\;t \in \mathcal{I}(k-1)$ be any one of the optimal solutions to \eqref{CBF generator}, with $x = x_0$ and $x = y_0$ respectively. We consider the difference between $B_k(x_0)$ and $B_k(y_0)$:
\begin{align}\label{a4}
	&B_k(x_0)- B_k(y_0) \nonumber\\
	=& \max\left\{\max _{t \in \mathcal{I}(k-1)} \alpha^{t} (h(x^*_{t} )+\lambda_t), \alpha^k B_0(x^*_{k} )\right\}-\max\left\{\max _{t \in \mathcal{I}(k-1)} \alpha^{t} (h(y^*_{t} )+\lambda_t), \alpha^k B_0(y^*_{k} )\right\}\nonumber\\
	=&\underbrace{\max\left\{\max _{t \in \mathcal{I}(k-1)} \alpha^{t} (h(x^*_{t} )+\lambda_t), \alpha^k B_0(x^*_{k} )\right\}-\max\left\{\max _{t \in \mathcal{I}(k-1)} \alpha^{t} (h(x'_{t} )+\lambda_t), \alpha^k B_0(x'_{k} )\right\}}_{\leq 0\;\mathrm{due}\;\mathrm{to}\;\mathrm{optimality}\;\mathrm{of} x^*_t}\nonumber\\
	&+ \max\left\{\max _{t \in \mathcal{I}(k-1)} \alpha^{t} (h(x'_{t} )+\lambda_t), \alpha^k B_0(x'_{k} )\right\}-\max\left\{\max _{t \in \mathcal{I}(k-1)} \alpha^{t} (h(y^*_{t} )+\lambda_t), \alpha^k B_0(y^*_{k} )\right\}\nonumber\\
	\leq&\max\left\{\max _{t \in \mathcal{I}(k-1)} \alpha^{t} (h(x'_{t} )+\lambda_t), \alpha^k B_0(x'_{k} )\right\}-\max\left\{\max _{t \in \mathcal{I}(k-1)} \alpha^{t} (h(y^*_{t} )+\lambda_t), \alpha^k B_0(y^*_{k} )\right\}\nonumber\\
	\leq&\max\left\{\max _{t \in \mathcal{I}(k-1)} \alpha^{t} h(x'_{t} )-(h(y^*_{t} )), \alpha^k (B_0(x'_{k})-B_0(y^*_{k} ))\right\},
\end{align}
where $x'_{t},\;t=0,1,...,k$ is the state trajectory starting from $x_0$ and applying $u^{y*}_t$, and the last inequality holds because of the triangle inequality for the maximum norm. Since (i) the trajectories $x'_{t}$ and $y^*_{t},\;t=0,1,...,k$ are obtained by applying the same control inputs, (ii) $f$ and $h$ are continuous in their domains, and (iii) the maximum of continuous functions of $x$ yields a continuous function of $x$, there exists a $\delta>0$ such that $B_k(x_0)- B_k(y_0) < \epsilon$ whenever $||x_0-y_0||_2 <\delta$. A mirror argument proves that $B_k(y_0)- B_k(x_0) < \epsilon$ whenever $||x_0-y_0||_2 <\delta$. The continuity of $B_k$ w.r.t. $x$ thus follows. 

Furthermore, if $f$, $h$, and $B_0$ are Lipschitz continuous, with Lipschitz constants $L_f$, $L_h$, and $L_{B_0}$, respectively, from \eqref{a4} we find that for any $x_0,\;y_0$ such that  $||x_0-y_0||_2 \leq\delta$, we have
\begin{equation}\label{a5}
	B_k(x_0)- B_k(y_0) \leq \underbrace{\max\left\{\max _{t \in \mathcal{I}(k-1)} \alpha^{t} L_h L^t_f, \alpha^k L_{B_0} L^k_f\right\}}_{:= L_{B_k}} \delta.
\end{equation}
Following a similar argument as for \eqref{a5}, we can get that $B_k(y_0)- B_k(x_0) \leq L_{B_k} \delta$ whenever $||x_0-y_0||_2 \leq \delta$. Thus, we conclude the Lipschitz continuity of $B_k$ w.r.t. $x$.

\subsection{Proof of Theorem \ref{theorem2}}\label{proof of Theorem2}
The proofs of (i) and (ii) can be combined. From Assumption \ref{A1} and \eqref{filter3}, we know that if \eqref{filter3} is recursively feasible for the initial state $x_0$, we have $Q(x_t, \pi_{\theta}(x_t)) \leq \Delta,\;\forall t\in \mathcal{I}(\infty)$, which further implies $B_k(x_t) \leq \Delta,\;\forall t\in \mathcal{I}(\infty)$. If $B_k$ is a CBF for \eqref{system} with the original state constraint $x\in X$, we get $h(x_t) \leq B_k(x_t) \leq \Delta$. Similarly, if $B_k$ is a CBF for \eqref{system} with the tightened state constraint $x\in X_\lambda$ and $\lambda \geq \Delta$, we get $h(x_t) \leq B_k(x_t)-\lambda \leq 0$. This completes the proofs of the statements (i) and (ii).

Next, we consider the last statement of Theorem \ref{theorem2}. For any initial state $x_0 \in \mathcal{S}_{B_k}$, there exists a $u_0$ such that $Q(x_0,u_0) \leq -\lambda$ according to \eqref{contractive} and \eqref{contractive_ex}. Since $Q_\theta$ satisfies Assumption \ref{A1}, $u_0$ makes $Q_\theta(x_0,u_0)\leq -\lambda + \Delta$, which means problem \eqref{filter3} is feasible at the initial state $x_0$. Then, we consider any feasible solution $u'_0$ to \eqref{filter3} when $x=x_0$, i.e., any $u'_0 \in U$ such that $Q_\theta(x_0,u'_0)\leq -\lambda + \Delta$. It follows from Assumption \ref{A1} and $\Delta \leq \lambda/2$ that $Q (x_0,u'_0) \leq -\lambda + 2\Delta \leq 0$. This means that $B_k(f(x_0,u'_0)) \leq 0$, i.e., the next state $x'_1 =f(x_0,u'_0)$ is in $\mathcal{S}_{B_k}$. As we have proved that problem \eqref{filter3} is feasible for any $x_0 \in \mathcal{S}_{B_k}$, the recursive feasibility of \eqref{filter3} is thus proved. A direct consequence of the recursive feasibility is that $B_k(x_t) \leq 0,\;\forall t\in \mathcal{I}(\infty)$, where $x_t$ is the state trajectory of the system $x_{t+1}=f\left(x_t, \pi_{\theta}(x_t)\right)$. Moreover, since $B_k$ is a CBF, we have $h(x_t) \leq 0,\;\forall t\in \mathcal{I}(\infty)$. This finishes the proof of (iii).
\subsection{Computing $B_0$}\label{computingB0}

In this subsection, we provide a detailed procedure of synthesizing the initial CBFs in Options 1-3. In Option 1, a CBF $B_0$ for the system \eqref{system} with the original state constraint needs to be obtained. In Option 2, a CBF for \eqref{system} with the tightened constraint $x \in X_\lambda$ is required. In Option 3, a contractive-set CBF is needed. 

We follow the similar design steps as presented in \cite{wabersich2022predictive,lazar2006stabilizing}. Firstly, we linearize the system \eqref{system} around the origin by $x_{k+1} = A x_k + B u_k + r(x_k,u_k)$, with the matrices $A:=\left.(\partial / \partial x) f(x, u)\right|_{(0,0)},\; B:=\left.(\partial / \partial u) f(x, u)\right|_{(0,0)}$, and the high-order error term $r(x(k), u(k))=f(x, u)-A x-B u$. Since $U$ is a polytope, we can compute its half-place representation $\{u\in \mathbb{R}^{n_u}| H_u x \leq h_u\}$. If the state constraint is linear, i,e., $h$ is the maximum of some affine functions, we compute the half-place representation of $X$ as $\{x\in \mathbb{R}^{n_x}| H_x x \leq h_x\}$. Otherwise, we compute a polyhedron $\{x\in \mathbb{R}^{n_x}| H_x x \leq h_x\}$ that is an inner approximation of $X$. To this end, we first specify $H_x$, which determines the shape of the polyhedron. The choice of $H_x$ is task-specific and the simplest choice is $H_x = [\mathbb{I}_{n_x}\;\;-\mathbb{I}_{n_x}]^T$, which will make the polyhedron a hyper-rectangle. Here, $\mathbb{I}_{n_x}$ denotes the $n_x\times n_x$ identity matrix. Then, we decrease $h_x$ until the following condition is verified:
\begin{align*}\label{verify1}
	\max_{x \in \mathbb{R}^{n_x}}\;\{h(x),\; \mathrm{s.t.}\; H_x x \leq h_x\} \leq \left\{\begin{aligned}
		0 \;&\mathrm{for} \; \mathrm{Options}\;1,3,\\ -\lambda \;&\mathrm{for} \; \mathrm{Option}\;2.
	\end{aligned}\right.
\end{align*}
A positive $h_x$ always exists if the origin is contained in $X$ for Options 1 and 3 or if the origin is contained in $X_\lambda$ for Option 2.

With the linearized system and constraints, we focus on finding a quadratic CBF of the form $B_0(x) = x^T E^{-1} x - \gamma$, where $E \in \mathbb{R}^{n_x \times n_x}$ is positive-definite and $\gamma$ can be taken as any positive value. To allow for solving convex optimization problems to obtain $E$, we parameterize a linear control law $u = Y E^{-1} x$, where $Y \in \mathbb{R}^{n_u \times n_x}$. As a result, the following lemma shows that one can make $B_0$ a (contractive) CBF for the linearized system by solving LMI inequalities. In the lemma, repeated blocks within symmetric matrices are replaced by $*$ for brevity and clarity.

\begin{lemma}\label{lemma3}
	Consider the CBF and control law parameterizations $B_0(x) = x^T E^{-1} x - \gamma$, $u(x) = Y E^{-1}x$. For the linearized system $x_{k+1} = A x_k + B u_k$ with the state and input constraints: $x \in \{x\in \mathbb{R}^{n_x}| H_x x \leq h_x\}$, $u \in U := \{u\in \mathbb{R}^{n_u}| H_u u \leq h_u\}$, we have the following results:
	
	(i) To make $B_0$ a CBF, it is sufficient to find $E$ and $Y$ such that 
	\begin{equation}\label{lmi1}
		\left[\begin{array}{cc}
			p E & EA^T+Y^T B^T \\
			* & E
		\end{array}\right] \succeq 0
	\end{equation}
	\begin{equation}\label{lmi2}
		\left[\begin{array}{cc}
			h^2_{u,i} & H_{u,i} Y \\
			* & E/\gamma
		\end{array}\right] \succeq 0,\; i=1,..., \dim(h_u)
	\end{equation}
	\begin{equation}\label{lmi3}
		\left[\begin{array}{cc}
			(h_{x,i}-q )^2 & H_{x,i} E \\
			* & E/\gamma
		\end{array}\right] \succeq 0,\; i=1,..., \dim(h_x)
	\end{equation}
	hold with $p=1$ and $q = 0$.
	
	(ii) To make $B_0$ a $\lambda$-contractive-set CBF satisfying \eqref{strenghten2} for a given positive integer $k$,  it is sufficient to find $E$ and $Y$ such that \eqref{lmi1}, \eqref{lmi2}, and \eqref{lmi3} hold with $p=1-\lambda/\gamma$ and $q=k\lambda \in (0, \min_{i} h_{x,i})$.
\end{lemma}
\emph{Proof:} By substituting the expression of $B_0$ and applying the Schur complement to \eqref{lmi1}, we get that $B_0$ satisfies $B_0(Ax+BYE^{-1}x) \leq p B_0(x) + p\gamma- \gamma,\;\forall x \in \mathcal{S}_{B_0}$. Applying the Schur complement to \eqref{lmi2}, we observe that \eqref{lmi2} is equivalent to $ YE^{-1}x \in U,\;\forall x \in \mathcal{S}_{B_0}$. Applying the Schur complement to \eqref{lmi3} yields that \eqref{lmi3} is equivalent to $H_{x,i} x - h_{x,i} + q \leq 0,\;\forall x \in \mathcal{S}_{B_0}$ and $\forall i \in \{1,...,\dim(h_x)\}$. By letting $p=1$ and $q = 0$, or letting $p=1-\lambda/\gamma$ and $q=k\lambda \in (0, \min_{i} h_{x,i})$, we prove the statements in Lemma \ref{lemma3}. \hfill$\square$

The condition $k\lambda \in (0, \min_{i} h_{x,i})$ indicates that the horizon $k$ should not be selected too large when using \eqref{CBF generator} to generate a contractive-set CBF.

Maximizing the volume of the safe set $\mathcal{S}_{B_0}$ is equivalent to maximizing the determinant of $E$. Therefore, we solve the following convex optimization problem:
\begin{align}\label{lmi4}
	&\min_{E,Y} \; -\log \operatorname{det}(E) \\
	&\mathrm{s.t.}\; \eqref{lmi1},\;\eqref{lmi2},\; \eqref{lmi3},\;\mathrm{and}\; E\succeq 0.\nonumber
\end{align}

After problem \eqref{lmi4} is solved, the validity of the obtained CBF $B_0$ for the nonlinear system \eqref{system} is verified via
\begin{align}\label{verify2}
	r \geq \max_{x \in \mathbb{R}^{n_x}}\;&f^T(x, Y E^{-1}x) E^{-1} f(x, Y E^{-1}x)-\gamma   \\
	\mathrm{s.t.}\; &x^T E^{-1} x - \gamma \leq 0 \nonumber,
\end{align}
where $r = 0$ for normal CBF verification and $r = -\lambda$ for $\lambda$-contractive-set CBF verification. In \eqref{verify2}, if the condition of \eqref{verify2} does not hold, we decrease $\gamma$ and repeat checking \eqref{verify2}. As shown in \cite[Section 2.5.5]{rawlings2017model}, there always exists a positive $\gamma$ such that \eqref{verify2} holds. 

\subsection{Detailed description of the case study}\label{case study details}
\begin{figure}\label{value_function}
	\centering
	\includegraphics[width=180pt,clip]{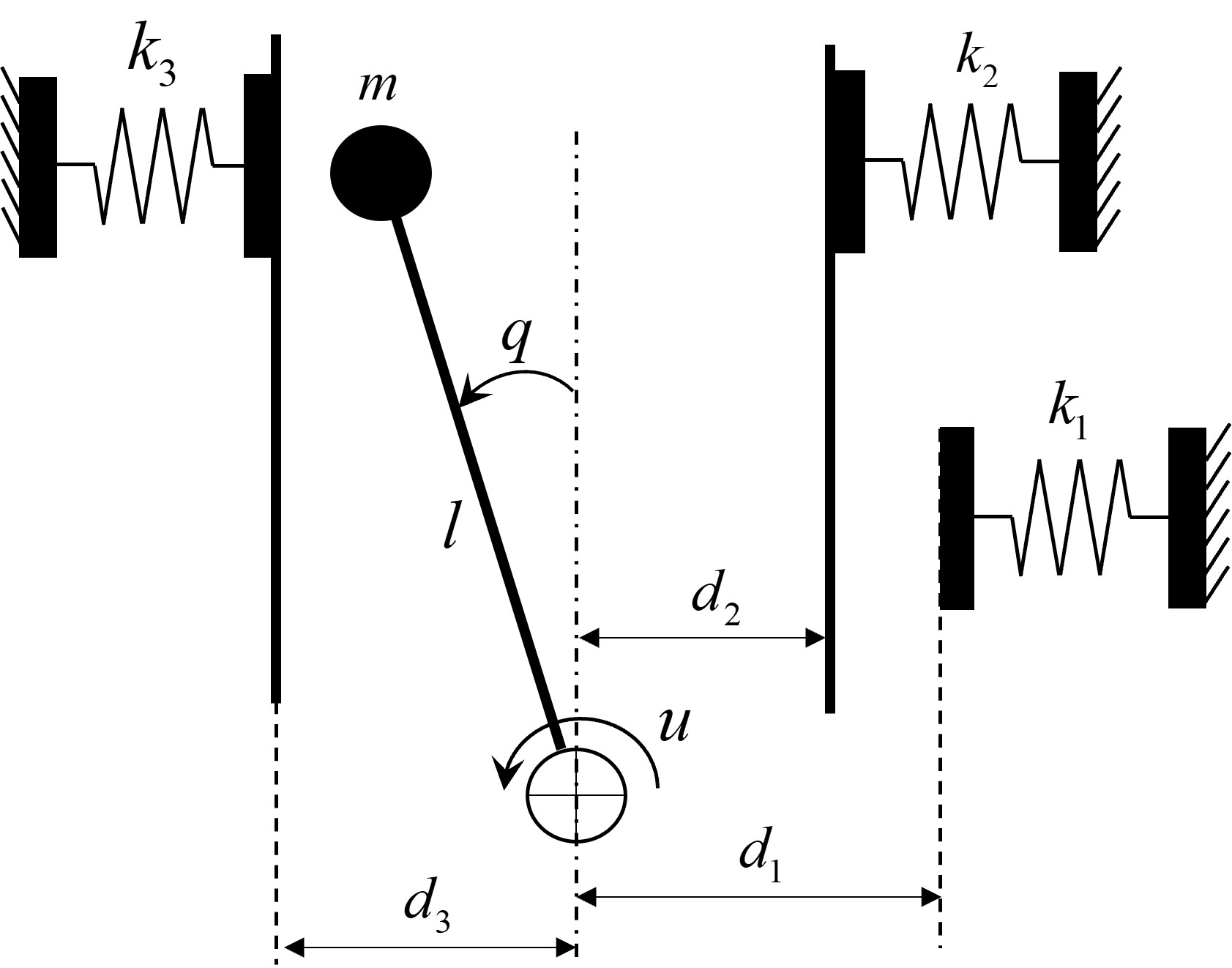}
	\caption{An inverted pendulum with elastic walls.}
\end{figure}

\textbf{Model:} Fig. 1 displays the inverted pendulum system. By linearizing the dynamics around the vertical configuration $\theta=\dot{\theta}=0$ and discretizing the system with a sampling time 0.05 s, we obtain a piecewise affine system:
\begin{equation*}
	x_{t+1}=A_i x_t+B u_t+f_i \quad \text { if }\; x_t \in \mathcal{C}_i,\;i=1,2,3,4,
\end{equation*}
where the system matrices and the regions are given by
\begin{align*}
	A_{1}&=\left[\begin{array}{cc}1 & 0.05 \\ -29.5 & 1\end{array}\right], \quad B=\left[\begin{array}{c}0 \\ 0.05\end{array}\right], \quad f_{1}=\left[\begin{array}{l}0 \\ -3.3\end{array}\right], \quad \mathcal{C}_1 = \{(x, u) \mid [1\;\;0]x \leq -0.12\},\\
	A_{2}&=\left[\begin{array}{cc}1 & 0.05 \\ -14.5 & 1\end{array}\right], \quad  f_{2}=\left[\begin{array}{l}0 \\ -1.5\end{array}\right],\quad  \mathcal{C}_2 = \{(x, u)\mid -0.12 \leq  [1\;\;0]x \leq -0.1\},\\
	A_{3}&=\left[\begin{array}{cc}1 & 0.05 \\ 0.5 & 1\end{array}\right], \quad  f_{3}=\left[\begin{array}{l}0 \\ 0\end{array}\right], \quad  \mathcal{C}_3 = \{(x, u)\mid -0.1 \leq   [1\;\;0]x \leq 0.1\},\\
	A_{4}&=\left[\begin{array}{cc}1 & 0.05 \\ -24.5& 1\end{array}\right], \quad  f_{4}=\left[\begin{array}{l}0 \\ 2.5\end{array}\right],\quad  \mathcal{C}_4 = \{(x, u) \mid [1\;\;0]x \geq 0.1\}.
\end{align*}

\textbf{Controllers:} We use four different controllers in the simulation. 

(i) MPC: We take the horizon $N=5$, 6, or 7. For the stage cost $C_\mathrm{MPC} : \mathcal{X} \times \mathcal{U} \to \mathbb{R}_{\geq0}$, we let $C(x,u)= x^T Q x + R u^2$ with $Q = \mathrm{diag}([20\;\;1])$ and $R=1$. The terminal stage cost is $x^T_N Q_\mathrm{T} x_N$ with $Q_\mathrm{T}$ obtained by solving the discrete algebraic Riccati equation for the subsystem $x_{t+1}=A_3 x_t+B u_t$ whose region contains the origin. The terminal constraint set is the maximal positively invariant set for the autonomous system $x_{t+1}=(A_3-BK) x_t$ with $K$ the solution to the above-mentioned discrete algebraic Riccati equation. Owning to the quadratic nature of the cost function and the PWA property of the predictive model, the MPC problem can be equivalently converted into a mixed-integer convex quadratic programming problem \cite{borrelli2017predictive}, and consequently, globally optimal solutions can be found efficiently by using the branch-and-bound approach \cite{wolsey1999integer}. We use Gurobi \cite{gurobi2021gurobi} to solve the MPC problem online. 

(ii) Supervised learning MPC: Following the approach in \cite{karg2020efficient}, we use an explicit controller, more precisely, an NN with three hidden ReLU layers containing 16, 32, and 8 neurons, to approximate the implicit MPC control law $u_\mathrm{MPC}$. In particular, we uniformly randomly sample the states from the state constraint set and select 4000 states that make the above-mentioned MPC problem with $N=7$ feasible. Next, 4000 state-input pairs $\{x_i,\;u_\mathrm{MPC}(x_i)\}_{i=1}^{4000}$ are obtained by solving the MPC problems with different initial conditions, and the NN is trained on these data pairs to approximate the map $u_\mathrm{MPC}$. 

(iii) Approximate dynamic programming: As demonstrated in \cite{he2023approximate}, ADP can produce a safe control policy by adding penalty terms in the cost function. In the ADP scheme, the objective is to find a control policy $\pi : \mathcal{X} \to \mathcal{U}$ to minimize the infinite-horizon cost
\begin{equation}\label{key}
	J_\pi\left(x_0\right)=\sum_{t=0}^{\infty} C_\mathrm{RL}\left(x_t, \pi\left(x_t\right)\right)
\end{equation}
for any initial state $x_0 \in \mathcal{X}$. Following the existing literature \cite{he2023approximate,beuchat2017approximate}, the state cost is specified as $C_\mathrm{RL}(x,u) = C_\mathrm{MPC}(x,u) + 30 \max\{0, |x_1| - 0.15\} + 3 \max \{0, |x_2| - 1\}$, where the second and third terms penalize the state constraint violation. The ADP algorithm uses a policy NN and a value NN, which evaluates $J_\pi$, to find the optimal policy iteratively, and terminates when the parameters of the value NN undergo sufficiently slight changes in two consecutive iterations. One can see Algorithm 2 in \cite{he2023approximate} for details. Although the training of the ADP policy accommodates state constraints, the ADP policy can still cause constraint violation because (i) the penalty terms in $C_\mathrm{RL}$ are external penalties and thus do not strictly guarantee constraint satisfaction, and (ii) there are approximation errors in the policy and value NNs. It can be observed from Table 1 that the proposed safe filters improve the safety of the ADP policy.

(iv) LQR: the LQR feedback control law $u_\mathrm{LQR}  = -Kx$, which is obtained when computing MPC in (i), can also be applied as the basic policy $\pi_0$ in the safety filters \eqref{filter} and \eqref{filter3}. When testing $u_\mathrm{LQR}$ without any safety filters, we simply project the resulting input value onto the input constraint set, which is a convex quadratic program. 

\end{document}